\documentclass[a4paper,12pt]{article}
\usepackage{amsmath,amssymb}
 \usepackage{epsfig} \usepackage{floatflt}
\usepackage{wrapfig}  \usepackage{subfigure}
 \graphicspath{{./figs14/}}
 \newcommand{\ns}{\normalsize}
 \newcommand{\fns}{\footnotesize}
\newcommand{\scs}{\scriptsize}
\newcommand{\beq}{\begin{equation}}
\newcommand{\eeq}{\end{equation}}                                  
 \def\eV{\relax\ifmmode{\rm e\kern-0.12em V}\else{\rm e\kern-0.12em V{ }}\fi}
\def\MeV{\relax\ifmmode{\rm M\eV}\else{\rm M\eV}\fi}
\def\GeV{\relax\ifmmode{\rm G}\eV\else{\rm G\eV}\fi}            
 \def\tildal{\relax\ifmmode{\tilde{\alpha}}\else{$\tilde{\alpha}${ }}\fi}
\def\agothk{\relax\ifmmode{\mathfrak A}_k\else{${\mathfrak A}_k${ }}\fi}
\def\tildals{\relax\ifmmode{\tilde{\alpha}(s)}\else{$\tilde{\alpha}(s)${ }}\fi}
 \def\al{\relax\ifmmode\alpha\else{$\alpha${ }}\fi}
\def\as{\relax\ifmmode \alpha_s\else{$ \alpha_s${ }}\fi}           
\def\alps{\relax\ifmmode\alpha_s\else{$\alpha_s${ }}\fi}
\def\msbar{\relax\ifmmode\overline{\rm MS}
\else{$\overline{\rm MS}${ }}\fi}
\def\albars{\relax\ifmmode{\bar{\alpha}_s}\else{$\bar{\alpha}_s${ }}\fi}
\def\albarsQ{\relax\ifmmode{\bar{\al}_s(Q^2)}\else{$\,\bar{\al}_s(Q^2)${}}\fi}
\def\agoth{\relax\ifmmode{\mathfrak a}\else{$\,{\mathfrak a}${ }}\fi}
 \def\acalkQ{\relax\ifmmode{\cal A}_k(Q^2)\else{${\cal A}_k(Q^2)${ }}\fi}
  \def\Acal{\relax\ifmmode{\cal A}\else{${\cal A}${ }}\fi}

\def\Orange  {}
 \def\myRed  {}
\def\Black{}

\begin{document}
 
 \centerline{\sf\Large A Few Lessons from pQCD 
 Analysis at Low Energies\footnote{The text of 
 contribution to Proceedings of ``Intern. Workshop 
 on $e^+e^-$ Collisions from $\phi$ to $J/\Psi$''
 (PHIPSI11), Novosibirsk Sept 2011; to be published
 in Nucl.Phys.(Proc.Suppl.)}} \smallskip
 
 \centerline{ D.V. Shirkov} \medskip 
 
 \centerline{\sf  Abstract} \smallskip 
  
 {\small Motivated by the recent 4-loop analysis of the 
 JLab data on Bjorken Sum Rule, where the pQCD series 
 seems to blow up at $\,|Q|\lesssim 1.5\,\GeV\,,\,\albars
 \gtrsim 0.33\,,$ we overview the general origin of the 
 divergency of common perturbation expansion over powers 
 of a small coupling parameter in QFT and consider in 
 detail the {\it blowing-up phenomenon} and accuracy of 
 finite sums for simple alternating and non-alternating 
 examples of divergent series.} 

 \section{ Introduction}
 It is known since the mid-XX that the main computational 
 tool of quantum theory, the perturbation expansion 
 $\sum_k\al^k\,c_k(...)\,$ over powers of the small 
 coupling parameter \al, is not a convergent one; 
 expansion coefficients grow factorially $c_k\sim k\,!\,.$ 
 The  reason is that every quantum amplitude (matrix 
 element) $C(\al, ...)$ is not a regular function of \al 
 at the origin $\al=0\,.$\\  
 \indent  Practically, the finite sum $\sum_k^N$ of such 
 a series could blow up at $N\sim1/\al\,.$ \\ To 
 illustrate, take a formal divergent series\vspace{-8mm}
  \begin{center}
\begin{tabular}{lr} \begin{minipage}{80mm}
 \beq\label{1} f(g)\sim\sum_{n\geq1}\,n !\,g^n
 \,= g + 2\,g^2\,+ \dots\,.\eeq \vspace{-5mm}
 
  Its finite sum \vspace{-2mm}
 \beq\label{2} f_{[k]}(g)=\sum_n^k\, f_n ;\quad 
 f_n=n !\,g^n\,, \eeq \vspace{-6mm}
 
 according to the Poincar\'e estimate \cite{poincare} 
 can approximate an expanded function $\,F\,$ with
 accuracy $\Delta_kf(g)= f(g)-f_{[k]}(g) \sim f_k\,.$
 \end{minipage}
&\hspace{2mm}\begin{minipage}{65mm}
\begin{center} \vspace{-6mm}
 $$\includegraphics[width=0.9\textwidth]
 {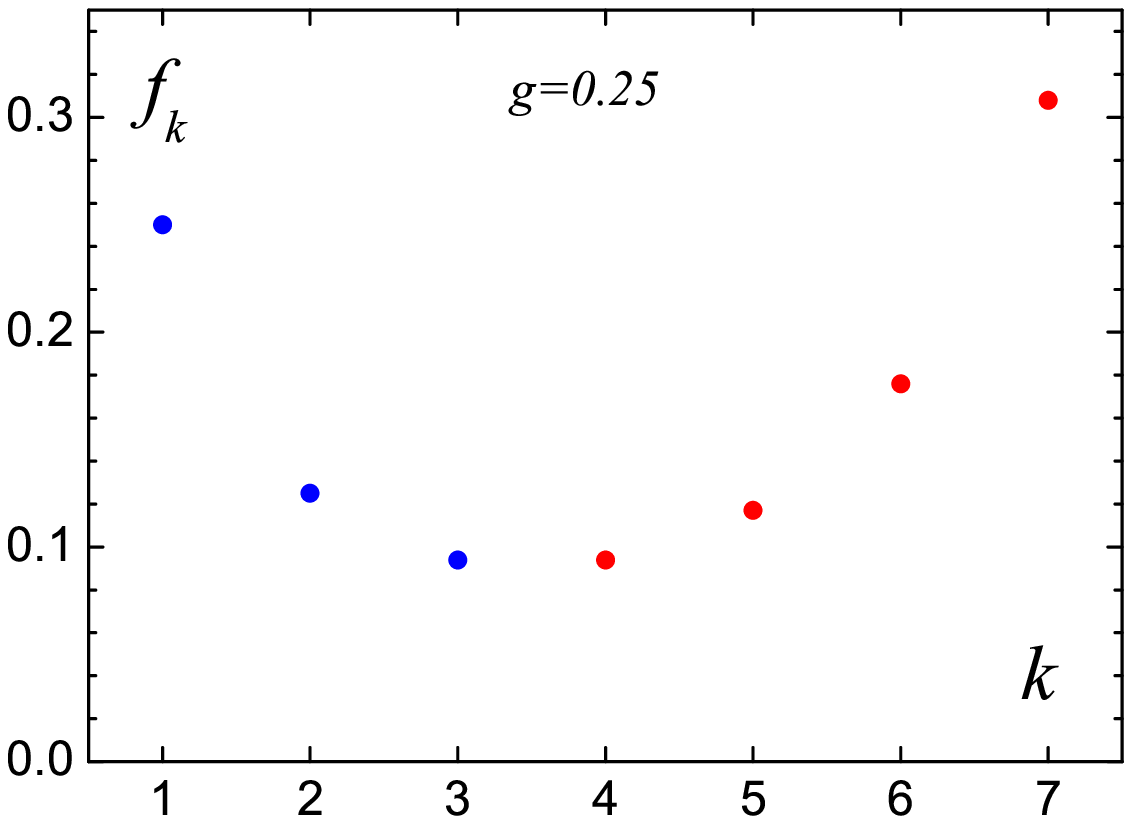}$$
  {\small Fig. 1  Values of $f_k$ terms at g=0.25.}
 \end{center} \end{minipage}
 \end{tabular} \end{center} 
 
 Thus, the finite sum can provide us with the best 
 possible accuracy \ $[\Delta_K f(g)]_{opt}=f_K\,$ at 
 an optimal number of terms 
 \beq k=K\sim 1/g\,.\eeq 
 The very existence of this \ {\sf lower limit of 
 possible accuracy} \ is an exact antithesis to the 
 case of  convergence series : any {\sf attempt to 
 increase} the number of terms above $K$ {\sf leads 
 to the lower accuracy}. At \ $g\lesssim1\,$ this 
 can happen for rather small $K\,$ values.
 \begin{quote}{\small\it In the above formal example (1), 
 at \,g=0.25, with K=4, $\,f_4(0.25)=0.5625\,$ and $\,
 \Delta_4F(0.25)=f_4=3/32$ this lower limit of accuracy 
 is about 16.7\,\%\,. For $\,f_5(0.25)=0.6798$, it is
 slightly worse -- 17.2 \%}.\end{quote}\smallskip 
 
 \section{Divergent Series and their Summation}
 \subsection{Explicit Illustrations} \  Consider 
 the integral\vspace{-1mm}
  \beq A(g)=\tfrac2{\sqrt{\pi}}\int^{\infty}_0\,
 e^{-x^2-(g/4)\,x^4}\,dx \,;\quad g > 0\,. \eeq

  Expanding integrand in $g$ and changing the order 
 of integration and summation one arrives at 
 alternating  divergent series 
 \beq A(g)=\sum_{n\geq0}\,(-g)^n\,A_n\,;\quad A_n=
 \frac2{4^n\sqrt{\pi}\,n\,!}\,\int\,e^{-x^2}\,x^{4n}
 \,dx \,;\quad A_0=1\,.\label{eq5}\eeq 
 The $n\to\,\infty$ limit for $A_n$ coefficients can 
 be estimated by the steepest descent method:  
 \[A_n^{as} \sim \int^{\infty}_0\,e^{n\,f(x)}\,dx\,,
 \quad  f(x)=4\,\ln x -\frac{x^2}{n}\,;\quad
 \mbox{with  result}\quad A_k^{as}=\frac{(k-1)\,!}
 {\sqrt{2\,\pi}}\,.\]
 
  Here, the divergent series was obtained by formal 
 manipulation with the finite expression. The finite 
 sums $a_{[n]}(g)=g\,A_1-\dots \pm A_n\,(-g)^n\,$ of 
 alternating series (\ref{eq5}) can be compared with 
 exact values of the function\footnote{Expressible 
 via  the particular Bessel function
 $ A(g)= e^{1/2g}\,(\pi\,g)^{-1/2}\,K_{1/4}(1/2g)$
 with known analytic properties. It is analytic in 
 the whole $g$ complex plane (cut along the negative 
 real semi-axis) with essential \ $\sim e^{-1/g}$ \ 
 singularity at the origin ; for details, see Sect.
 2.2 in paper \cite{KazSh80}.} $A(g)=1-a(g)\,.$ 
 For results of comparison see Fig.\ref{fig:a[n]}. 
 There, we show that starting from $g=0.25\,$ the 
 $a_{[4]}$ curve passes farther from the exact one 
 than the $a_{[3]}$ curve. 
  \addtocounter{figure}{1}
  \begin{figure}[h!] 
 \centerline{\epsfig{file=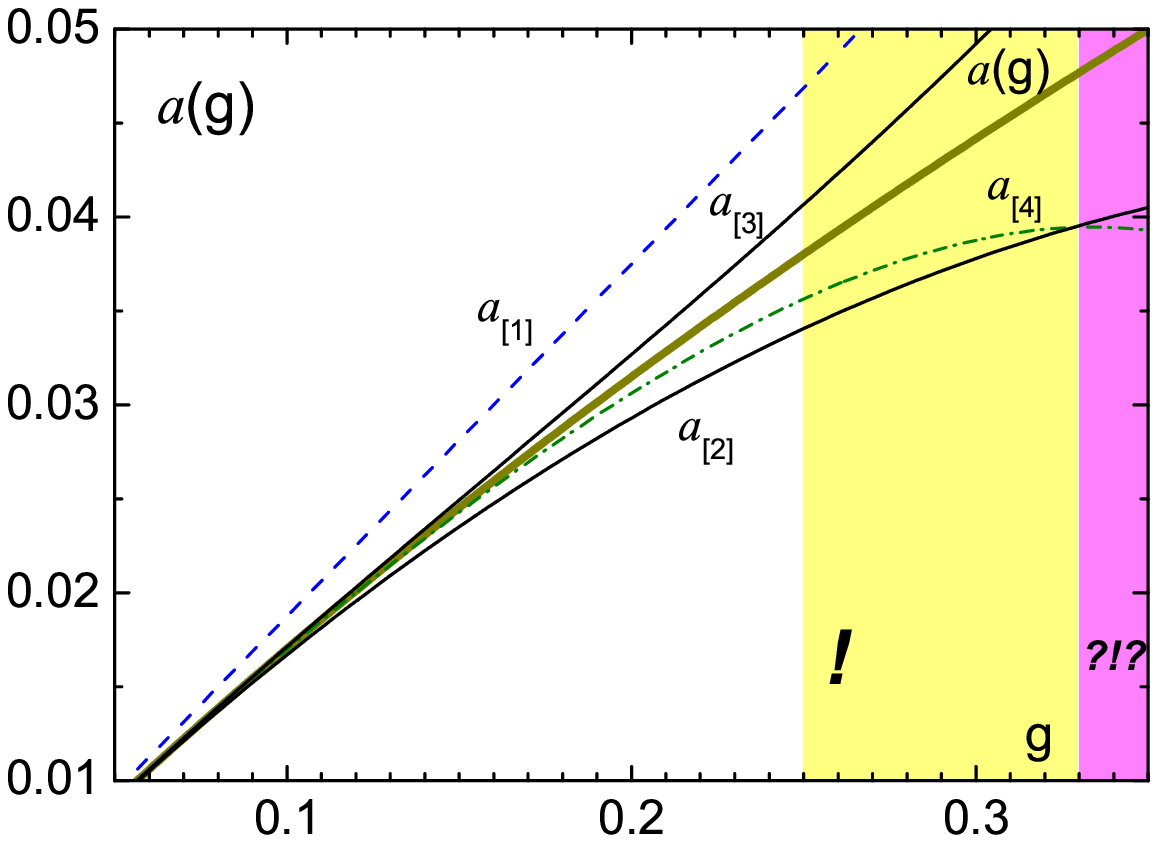,width=8.6cm}} 
 \vspace{-8mm}
 \caption{\fns The $a_{[k]}$ approximants for the 
 function $A(g)$. The mark of exclamation, ``!'' 
 denotes the beginning of the yellow zone (= caution 
 light) while the combination ``?!?''  marks 
 the red zone.} \label{fig:a[n]}\end{figure}
 
  A practical example of alternating divergent 
 series gives the beta-function of the $g\varphi^4$ 
 model. In the late 70s its expression 
  $$\beta_{\msbar}^{pt,4}(g)=\tfrac3 2\,
 g^2-\tfrac{17}{6}g^3+16.27\,g^4 - 135.8\,g^5 $$
 known up to the $N^3LO$ term was used \cite{KazSh80} 
 as a starting point for the whole function
 $\,\beta_{\msbar}(g)\,$ restoration. In the 
 reconstruction procedure (based also on asymptotic
 expression \cite{lip77} for $\,\beta_n^{as}\,$) the 
 Borel representation 
 supplemented by conformal transformation was 
 involved. The resulting\footnote{with $b_3(x)$ 
 being the cubical polynomial in $w(x)\,,$ the 
 conformal variable.} closed formula 
 \beq\beta_{\msbar}^{\scs CB}(g)=\int^{\infty}_0\,
 \frac{dx}{g} e^{-x/g}\left(\frac{d}{dx}\right)^5
 \,x^2\,b_3(x)\,\eeq 
 can be used for next coefficient estimation. 
 Later on, the next $N^4LO$  term was calculated
 \cite{beta5k,beta5ch} $\beta_5=1420.6\,$ via 
 Feynman diagrams. Comparing it with the prediction
 (6) $\beta_5^{CB}=1409.6\,$ \ gives the accuracy
 within 1 \% ! \vspace{4mm}

\begin{figure}[h!]
 \centerline{\epsfig{file=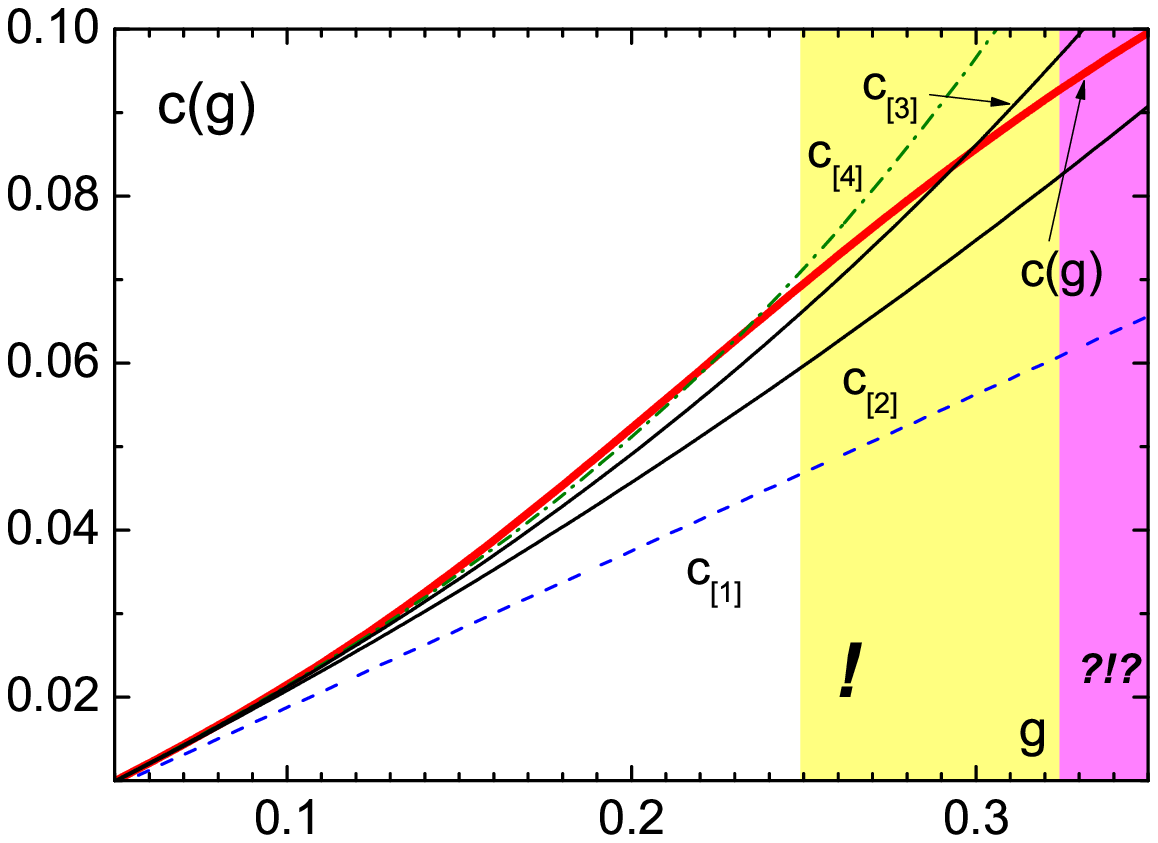,width=8.6cm}}
 \vspace{-8mm} 
 \caption{\fns The $c_{[k]}$ approximants for the 
 function $c(g)=C(g)-1\,.$}  \label{fig:c[n]}\end{figure}

  Another model integral 
 \beq C(g)=\tfrac1{\sqrt{\pi}}\int^{\infty}_{-\infty}
 \,e^{-x^2(1-\frac{\sqrt{g}}4\,x)^2}\,dx \to \sum_k 
 g^k C_k\,;\quad C_k=A_k=\left.\tfrac{\Gamma(2k+1/2)} 
 {4^k\,\Gamma(k+1)}\right|_{k\gg1} \to \,
 \tfrac{\Gamma(k)}{\sqrt{2\,\pi}}\, \eeq
 produces non-alternating asymptotic power series with 
 the \underline{same} coefficients. As far as this 
 integral is also expressible in terms of Bessel 
 functions (see Ref.\cite{KazSh80}, page 482), one has 
 exact expression for the coefficients and can compare 
 the finite sum approximations
 $c_{[n]}(g)=C_1+\dots+C_n\,g^n\,$ with exact values 
 of $C(g)-1=c(g)\,$ -- see Fig.\,\ref{fig:c[n]}. 
 
 It is clear that the 2-term approximant (lower thin 
 curve) is good up only to $g=0.15-0.20$ and the 3-term 
 one (upper thin curve) up to $g\sim 0.33\,$ while the 
 4-term sum (upper broken curve) starts to deviate from 
 $C(g)$ (red thick curve) at $g\sim 0.27\,!$ \medskip
 
  The model (7) is more instructive for our case
 motivated by the fresh signal from the perturbative 
 Quantum Chromodynamics (pQCD) in the low-energy domain. 
 There, the 4-loop analysis of rather precise JLab data 
 on polarized Bjorken Sum Rule revealed \cite{BjSR-11} 
 that the non-alternating series for the pQCD 
 correction (eq.(3) in \cite{BjSR-11})
 \beq\label{bjork} \Delta_{[4]}^{Bj}(\as)=0.3183\,\as
 +0.3631\,\as^2 + 0.6520\,\as^3 + 1.804\,\as^4 \eeq
 does blow up at $\,\as \sim 0.35\,.$ It is noteworthy
 that the coefficient ratios here (1.1,\,1.8,\,2.8) 
 are close to the factorial ones (1,\,2,\,3).
 
  \begin{figure}[h!]
 \centerline{\epsfig{file=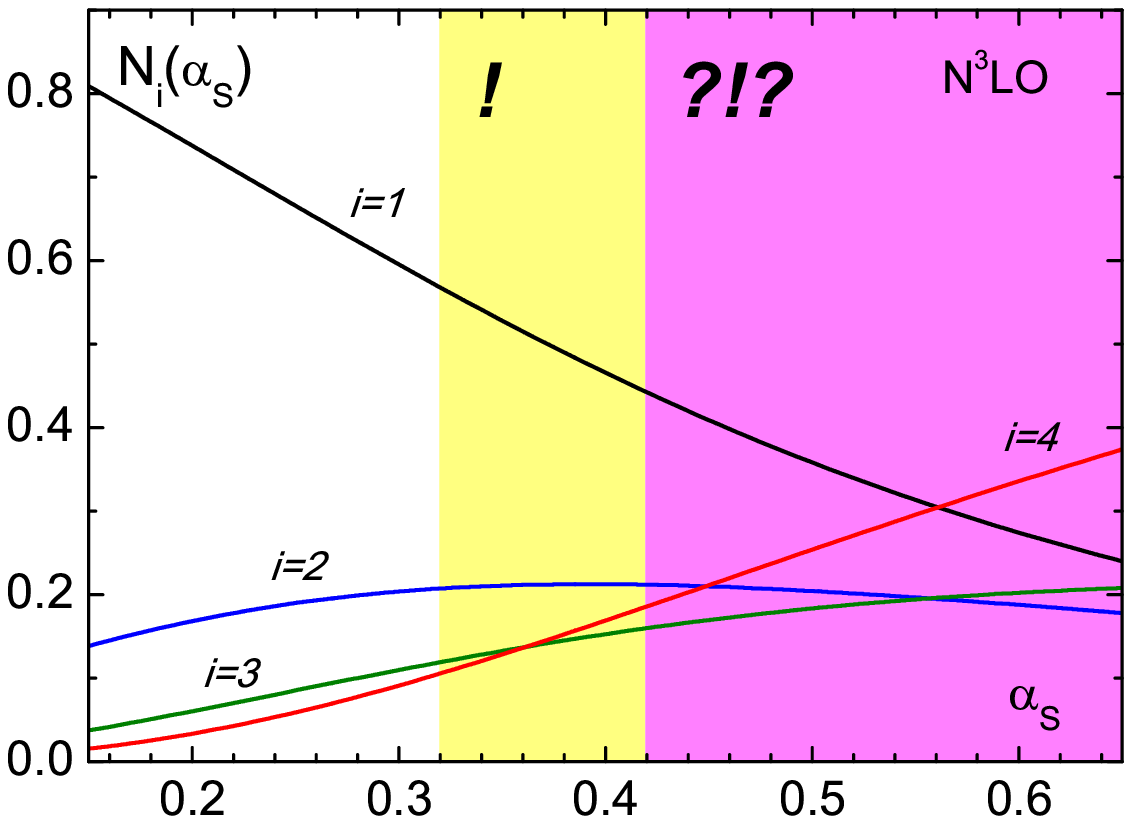,width=7.6cm}}
 \vspace{-6mm}
 \caption{\fns The \as-dependence of the relative PT 
 contributions to the Bjorken amplitude at the 4-loop 
 level -- based on Eq.(9).}\label{bjPT}\end{figure} 

 Indeed, as it is shown on Fig.\ref{bjPT}, the 4-loop 
 term ($\sim\as^4$) is close to the 3-loop one in the 
 interval $ 0.3\lesssim\as\lesssim 0.4\,,$ while it 
 approaches the 2-loop term at $\as \geq 0.4\,.$ We 
 marked the first region as a ``yellow zone'' \ and the 
 second as a ``red'' one. Roughly, this corresponds 
 to the rule $K\sim 1/\as\,,$ Eq.(3) with $K=i-1\,.$\\
 
   Two other illustrations on Fig. 5 a,b,  also taken 
 from paper \cite{BjSR-11} demonstrate the lack of 
 progress in the 4-loop description - marked by black 
 hatching (SW-NE direction)- with respect to the 
 3-loop one (red hatching in the NW-SE direction).
 
 \begin{tabular}{lr}
 \begin{minipage}{70mm}
 $$\includegraphics[width=\textwidth]
 {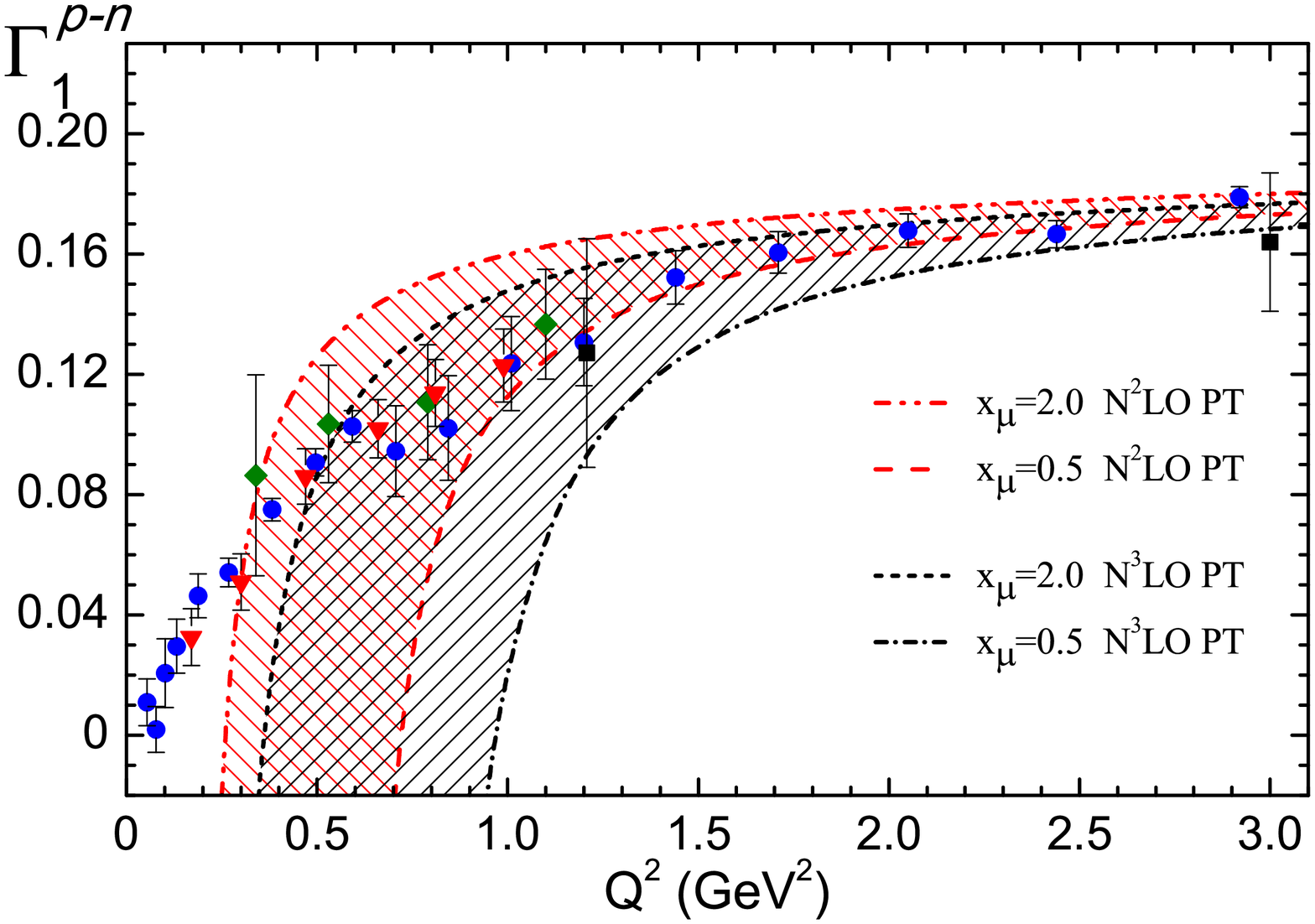}$$  \vspace{-11mm}
 
 {\fns Fig. 5a \ The QCD perturbation analysis of \\
 the \ Bjorken form-factor \ confronted with \\ 
 JLab data in three- and four-loop orders.} 
 \end{minipage}
 &\hspace{-2mm}\begin{minipage}{72mm}
 \begin{center} 
 $$\includegraphics[width=1.04\textwidth]
 {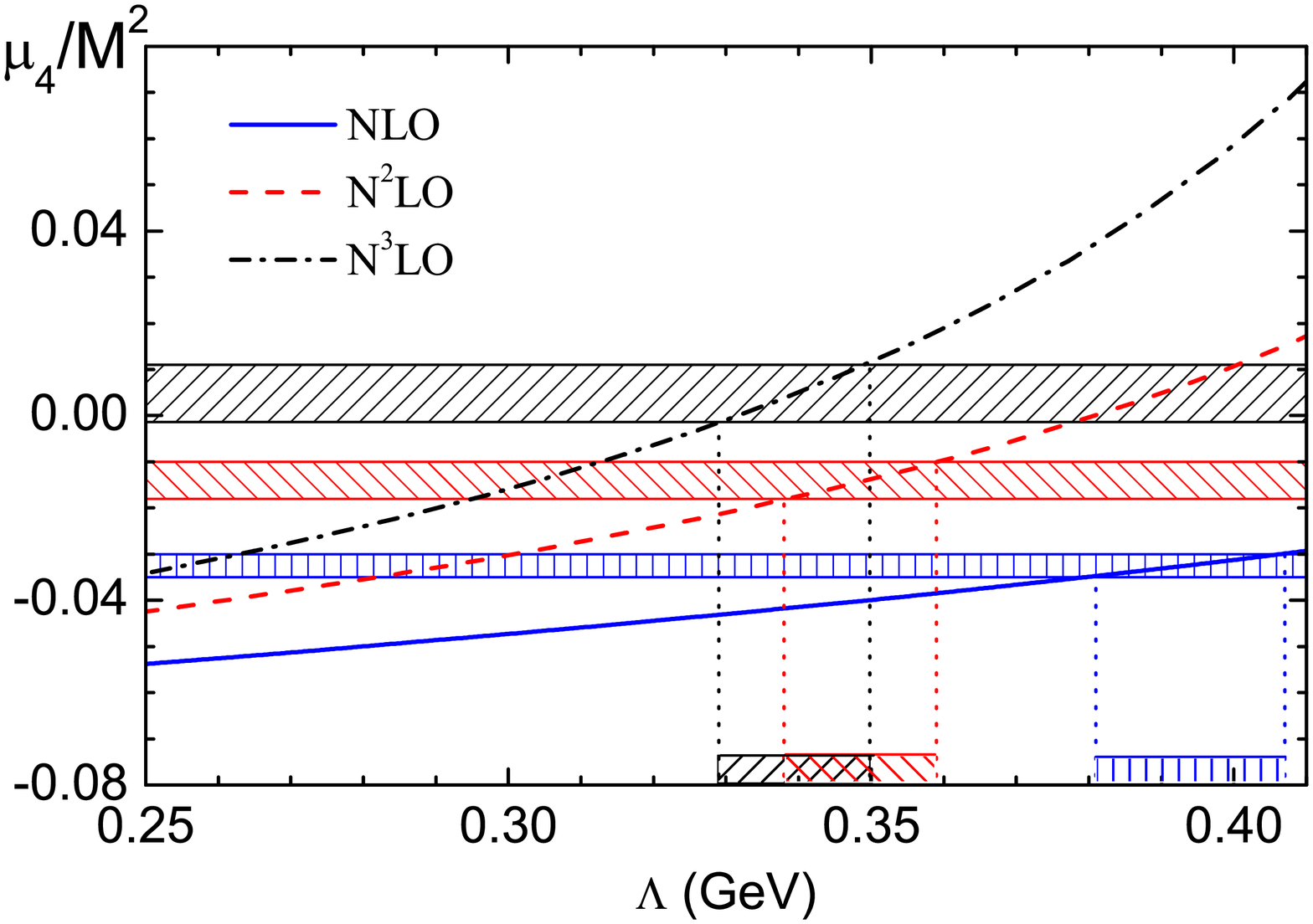}$$  \vspace{-10mm}
 
 {\fns Fig. 5b \ Instability of the HT coefficient 
 $\mu_4$ fitting of the Jlab data, as in Fig.5a 
 } \end{center} \end{minipage}
 \end{tabular} 
 
 \subsection{Asymptotic series and essential  
   singularity} 
 Turn to the origin of the non-convergent asymptotic
 series (AS) like in Eqs.(1),(6),(8). Usually, it is 
 related with the essential singularity at the origin 
 $\,\al=0\,$ that is a common property (in the 
 theories of Big Systems) of the objects representable 
 via Functional or Path Integral. This is the case for 
 Turbulence, Classic and Quantum Statistics and 
 Quantum Fields. Numerous examples are well known : 
 the $\,\mathbf e^{-1/g}\,$ dependence of the energy 
 gap in BCS and Bogoliubov theories of 
 SuperConductivity : for tunneling probability in 
 quantum mechanics. In the theory of Quantum fields 
 (QFT) it was first discussed for QED by Dyson just 
 60 years ago \cite{Dyson52} and soon after that 
 implemented by Bogoliubov \cite{NN+aa+dv}; (for the 
 QCD case, the same method was used in the so-called
 APT approach -- see below Section 3).
 
 Mathematically, the essential singularity origin is 
 connected with the small parameter $g\,$ (or \al) 
 attached to some nonlinear structure. In the quantum 
 case, this is interaction term. Generally, a certain 
 AS corresponds to a set of various functions. Hence, 
 in physics, \\ \centerline{\sf The Asymptotic Series 
 \ ``summation'' \ is an Art.}\smallskip
  
  This motto really implies that for the adequate AS
 summation one should involve some additional 
 arguments, like in the Eq. (6) example above. 
 
 \subsection{ Higher PT terms for \           
  $\mathbf{e^+e^-\to}$ hadrons} \vspace{-2mm}
 As far as this meeting is devoted mainly to the
 electron-positron collider physics, turn to the 
 inclusive $e^+e^-\to$ hadrons process. Two functions,
 the cross-section ratio $R(s)=1+r(s)\,,$ and the Adler
 function $D(Q^2)=1+d(Q)\,$ are in use there. Table 1
 presents the short summary of the PT terms relative 
 contribution in the `moderate energy' \ interval 
 below $m_\tau\,.$ 

 \begin{center}{\sf\ns Table 1. Relative contributions 
 \ of 1-, ... \ 4-loop terms in \ $e^+e^-\to$ \ hadrons}
 \smallskip 

 \begin{tabular}{|c|c||c|c|c|c|} \hline
 {\scs Function}&{\tiny Scale/Gev}&\multicolumn{4}{c||}
 {\small PT terms \ (in \%)} \\ \hline\hline
 \multicolumn{2}{|c|}{\scs the loop number 
 $\ell \ \to$} &1&2&3&4 \\ \hline\hline
 r(s)&1& \ 65 \ & \ 19 \ &{\bf 55 ?!?}&{\bf -39 ?!?} 
  \\ \hline
 r(s)&1.78&73&13&{\bf24 ?!?}&{\bf-10 ?!?}\\ \hline\hline
 d(Q)&1&56&17&{\bf\Orange11 !}&{\bf\myRed 16 ?!?}
 \Black \\ \hline
  d(Q)&1.78& 75 &14& 6 &{\bf\Orange 5 !}\Black \\ \hline 
  \end{tabular} \end{center}\smallskip  
   
 \ns In the upper two lines, for $r(s)\,,$ one can 
 see the literally terrible effect of the $\pi^2$ terms 
 on the higher $\ell=3,4\,$ contributions. This issue 
 was resolved in the 80s\cite{Rad82,KraPivo82}. 
 The net result is that in the annihilation channel, 
 the s-channel, one should use some special QCD coupling 
 $\tildal(s)\,$ instead of $\albars(Q^2)\,.$ See below, 
 eq.(9) and Fig.6a.
 
 Concerning the higher contributions, 
 $d_{\ell=3,4}\,,$  to the Adler function one 
 observes the picture analogous\footnote{
 with due account of the QCD common coupling values 
 $\,\albars(1\,\GeV)=0.55\,$ and $\,\albars(m_{\tau})= 
 0.35\,.$} to the one illustrated by Fig.4.
  
 \section{\ns Analytic Perturbation Theory}         
 \subsection{\small A Few Words about APT} 
 Analytic Perturbation Theory (APT) in QCD, is the 
 closed theoretical scheme devised\footnote{See also 
 review papers\cite{ShSol99,Sh06,ShSol06}.} in the 
 mid-90s \cite{SolSh90s} without
 Landau singularities and additional parameters. It 
 stems from the imperatives of \ RG-invariance, 
 $Q^2$-analyticity, compatibility with linear integral 
 (like, the Fourier) transformations and essentially
 incorporates non-perturbative $\,e^{-1/\as}$ 
 (algebraic in $Q^2$)\footnote{For the deep connection 
 between the \al-non-perturbativity and the 
 $Q^2$-analyticity, see Ref.\cite{sh76}} structures.
 \smallskip

  Instead of the power PT set  \albarsQ,\albarsQ$^2$,
 \albarsQ$^3,\dots$ one has a non-power APT expansion 
 set \ \{\acalkQ\} $k=1,2,\dots$  with all  \acalkQ 
 regular in the IR region. Accordingly, for the 
 s-channel, there is another IR-regular set 
 $\tildal_k(s)\,.$ The first functions $\Acal_1(Q^2)=
  \al_{an}\,,$ and $\tildal_1(s)=\tildal(s)$
 at the one-loop case look rather simple
 \beq \al_{an}(Q^2)=\frac1{\beta_0\,\ln(Q^2/\Lambda^2)} 
 + \frac{\Lambda^2}{\beta_0\,(\Lambda^2-Q^2)}\,;\quad
 \tildal(s)=\frac1{\pi\,\beta_0}\,
 \arctan\frac{\pi}{\ln(Q^2/\Lambda^2)}\,.\eeq
 Both are presented on Fig.6a \ together with common 
 \albars, singular at $Q=\Lambda=400\,\MeV\,.$ Their 
 regular LE behavior corresponds quantitatively to 
 results of lattice simulation (see Fig.6b) down to 
 $Q\sim 500\,\MeV.$\vspace{-4mm}
 
  \begin{center}
\begin{tabular}{lr}
 \begin{minipage}{75mm}
 \begin{center} \vspace{-6mm}
 $$\includegraphics[width=0.9\textwidth]
 {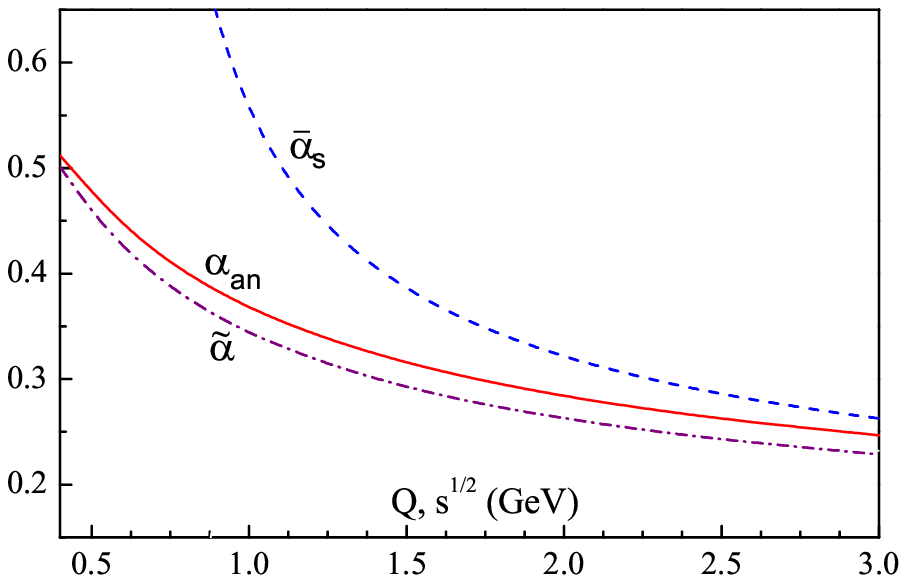}$$
 {\fns Fig.6a \ Analytic QCD couplings $\al_{an}(Q)$ 
 and $\tildal(s)$ in comparison with common \albars.}
 \end{center} 
\end{minipage}
&\hspace{2mm}\begin{minipage}{70mm}
\begin{center}  \vspace{-3mm}
 $$\includegraphics[width=0.8\textwidth]{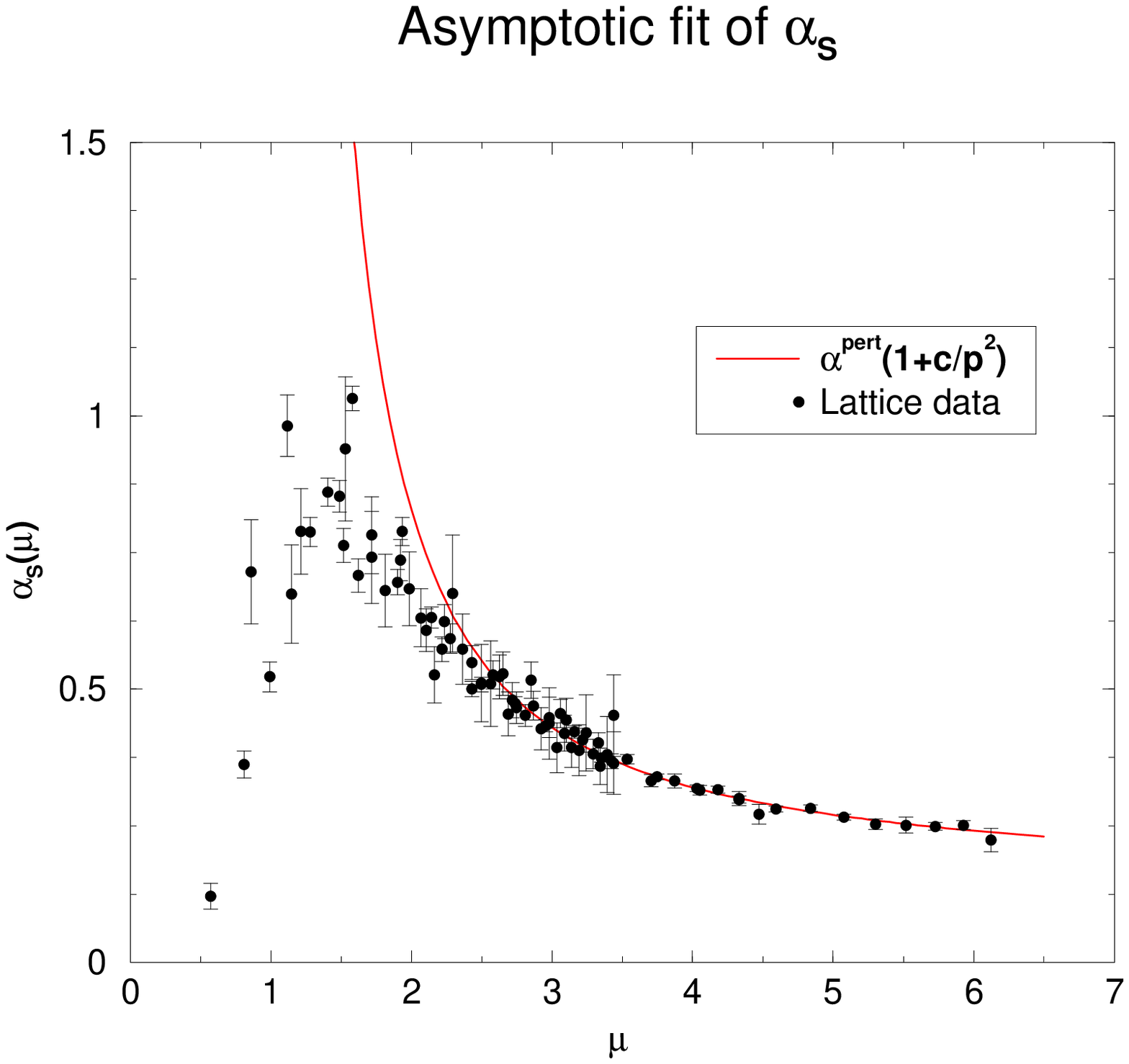}$$
  \vspace{-11mm}
  
 {\fns Fig.6b \ The lattice \as based on three-gluon 
 vertex}
 \end{center} \end{minipage}
 \end{tabular} \end{center} \bigskip

\ns As it can be seen from Fig.7 the APT+Higher Twist
 (HT) description of the JLab data looks quite 
 satisfactory down to 350-400 \MeV, that is to 
 the $\Lambda_{QCD}$ scale !
 
 \addtocounter{figure}{2}
  \begin{figure}[h!]
 \centerline{\epsfig{file=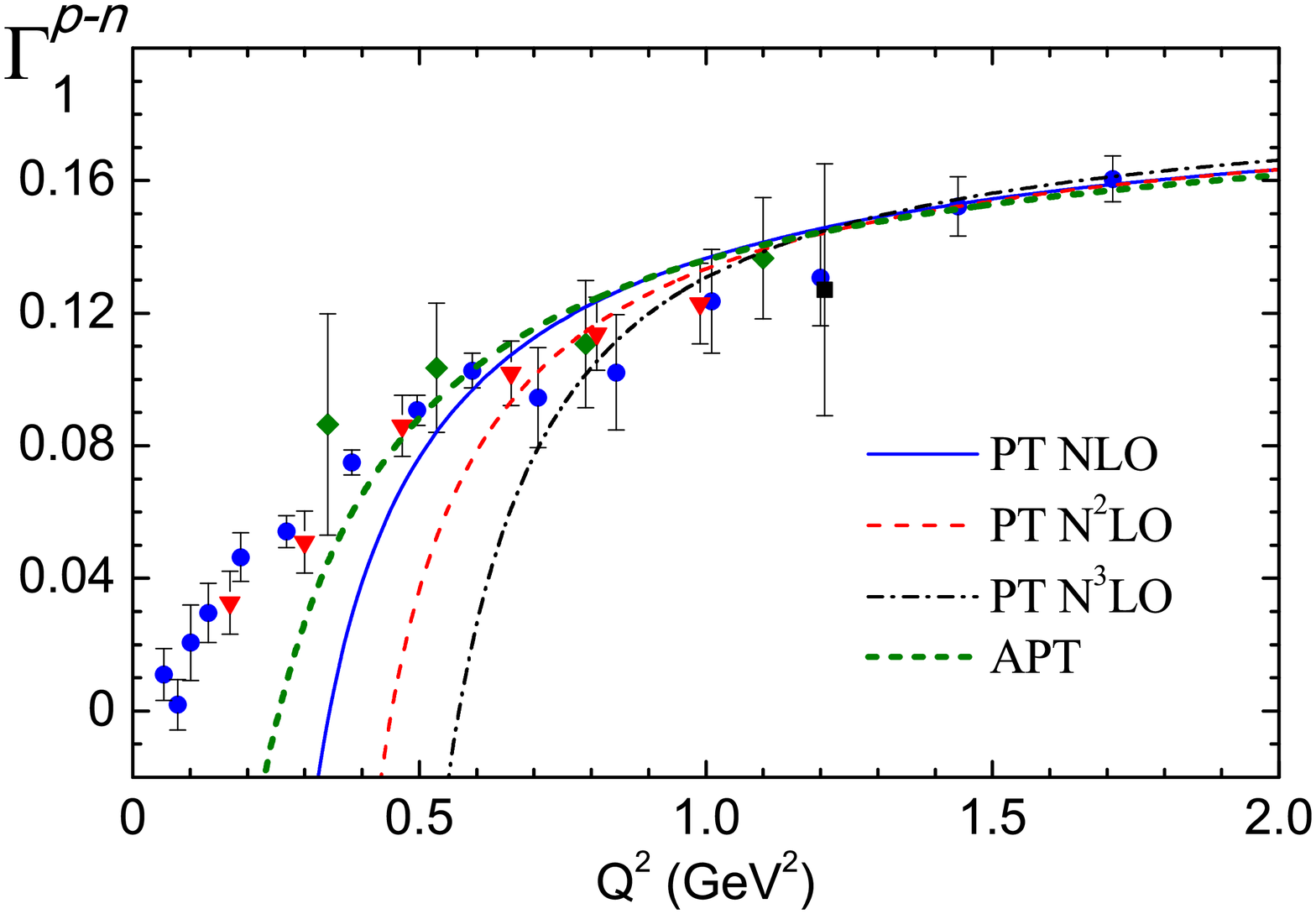,width=9.6cm}}
 \vspace{-4mm}
 \caption{\fns Reasonable fit of the JLab data by 
 APT supplemented by Higher Twist (HT) terms (the 
 upper green dotted curve) down to 350 \,\MeV\,. 
 Three lower curves describe the standard PT 
 fits.} \label{apt fit}\end{figure} 

  We omit here technical details of the APT+HT
 analysis of paper \cite{BjSR-11}. Some of them can 
 be seen in the last right columns of Table 2.
 There, higher PT and APT contributions to couple 
 of sum rules are summarized.\smallskip

 \begin{center}
 {\sf Table 2. \ Relative contributions (in \%) of 
 \ 1-,2-,3- and \ 4-loop \ terms}\small

\begin{tabular}{|c|c|c||c|c|c|c||c|c|c|} \hline
 \multicolumn{2}{|c|}{\slshape Process}& {\scs
 Scale/Gev}&\multicolumn{4}{c||}{\slshape PT(in \%)}
 & \multicolumn{3}{r|}{\slshape APT ${^*}$
 \phantom{aa}} \\    \hline 
 \multicolumn{3}{|c|}{the loop number =}&1&2&3&4& 
 1 &2 &3   \\    \hline\hline
  Bjorken SR&t&1&35&20&{\bf\Orange19 !}&
  {\bf\myRed26 ?!?}& \Black 80 & 19 &{1} \\ \hline
  Bjorken SR&t & 1.78 &56 &21 &13&{\bf\Orange11 !}
  \Black &80 & 19 &{1} \\ \hline
 {\fns GLS SumRule}&t&1.78&65 &24 &{\ns\bf\Orange 11 
 !}& \Black &75 & 21 & { 4}  \\ \hline
 {\fns Incl. $\tau$-decay}&s&1.78 &51 & 27 & 14 &
 {\bf\Orange 7 !}\Black& 88 & 11 &{1} \\ \hline
 \end{tabular} \smallskip 
 
 * {\small The 4-loop APT contributions are 
 negligible everywhere.} \end{center}
 
 \centerline{\bf\ns Invitation for Work 
 \small(instead of Conclusion)}\smallskip
 
  A number of topics is in order: \vspace{-4mm}
  
  \begin{itemize} \itemsep -1.51mm 
 \item Devising methods of AS summation, 
 (including integral and conformal tricks),
 \item Devising Generating Function for HT terms in QCD
  \item either generalizing the minimal APT, 
 \item  Toy models for the 4-loop term predicting
   for other processes \ $P_i$
 \item Set of analytic couplings $\,\alpha_s^i\,,$
     each being adequate to a given process \ $P_i$ \ ?
 \item Generating \ HT \ function for the each \ $P_i$ \ ?
 \end{itemize}
 
  \centerline{\ns\sf Acknowledgments} \medskip
 It is a pleasure to thank Dr. V. Khandramai for useful 
 discussion and technical help. This research has 
 partially been supported by the presidential grant 
 Scient. School--3810.2010.2, RFFI grant 11-01-00182  
 and by the BelRFFR-–JINR grant F10D-001.\small 

  \end{document}